\documentclass[manuscript,screen,anonymous=false, nonacm]{acmart}

\AtBeginDocument{%
  \providecommand\BibTeX{{%
    \normalfont B\kern-0.5em{\scshape i\kern-0.25em b}\kern-0.8em\TeX}}}

\usepackage{subcaption}
\usepackage{tabularx}
\usepackage{xcolor, listings, framed}
\usepackage{xspace}

\definecolor{shadecolor}{gray}{0.95}
\newcommand{\system}{Keyframer\xspace}

\newcommand{\rqone}{What painpoints exist for motion designers, and what ideas do they have for how AI might assist with these processes?}
\newcommand{\rqtwo}{What design strategies do users take to prompt LLMs for animations using natural language?}
\newcommand{\rqthree}{How does \system support iteration in animation design?}
\begin{document}

\title{Keyframer: Empowering Animation Design Using Large Language Models}

\author{Tiffany Tseng}
\email{tiffanytseng@apple.com}
\affiliation{%
  \institution{Apple}
  \country{USA}
}

\author{Ruijia Cheng}
\email{rcheng23@apple.com}
\affiliation{%
  \institution{Apple}
  \country{USA}
}

\author{Jeffrey Nichols}
\email{jwnichols@apple.com}
\affiliation{%
  \institution{Apple}
  \country{USA}
}

\begin{abstract}
Creating 2D animations is a complex, iterative process requiring continuous adjustments to movement, timing, and coordination of multiple elements within a scene. To support designers of varying levels of experience with animation design and implementation, we developed Keyframer, a design tool that generates animation code in response to natural language prompts, enabling users to preview rendered animations inline and edit them directly through provided editors. Through a user study with 13 novices and experts in animation design and programming, we contribute 1) a categorization of semantic prompt types for describing motion and identification of a ‘decomposed’ prompting style where users continually adapt their goals in response to generated output; and 2) design insights on supporting iterative refinement of animations through the combination of direct editing and natural language interfaces.
\end{abstract}



\begin{teaserfigure}
  \includegraphics[width=\textwidth]{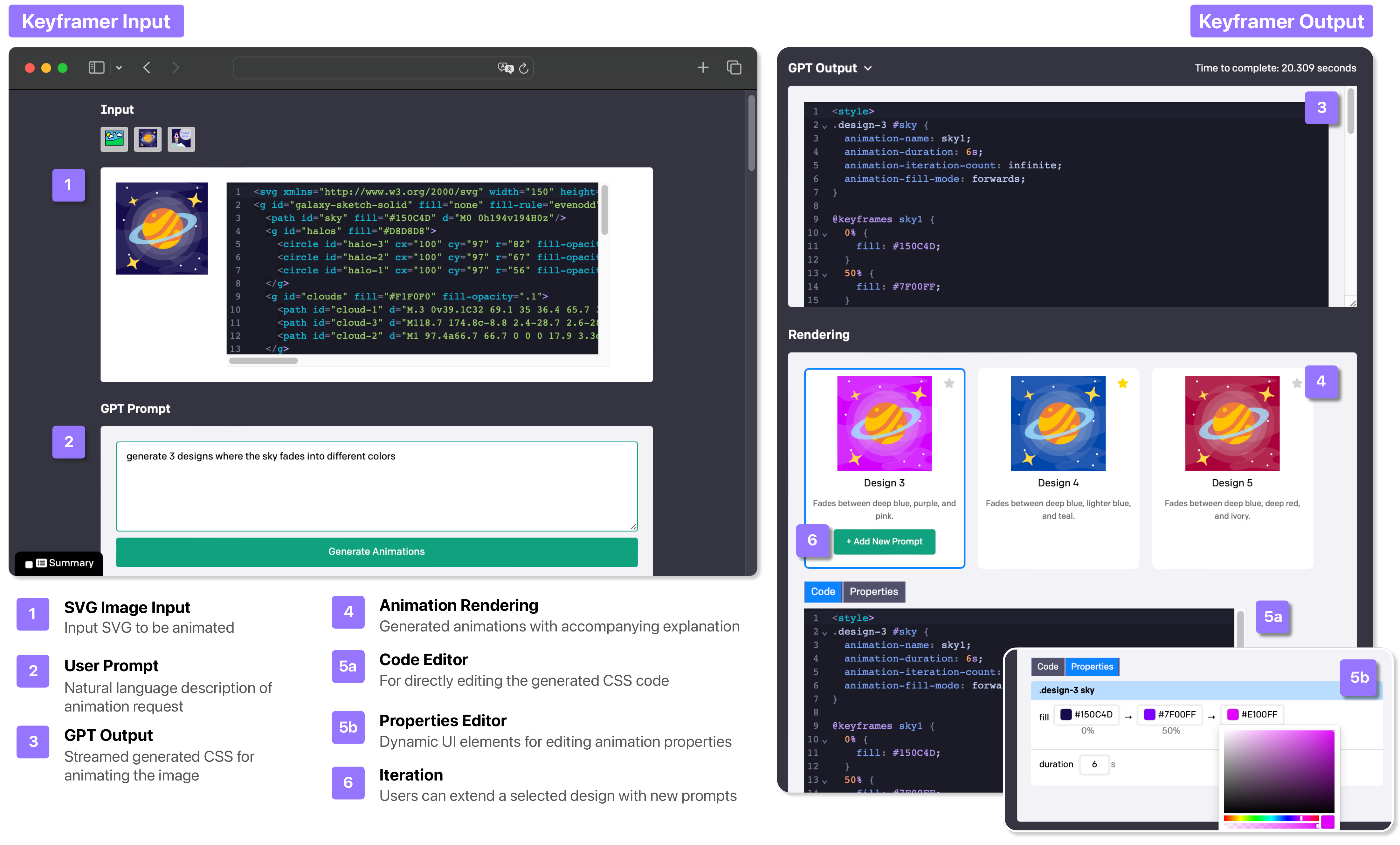}
  \caption{Keyframer is an LLM-powered animation prototyping tool that can generate animations from static images (SVGs). Users can iterate on their design by adding prompts and editing LLM-generated CSS animation code or properties. Additionally, users can request design variants to support their ideation and exploration.}
  \Description{Keyframer enables users to generate animations from static images, leveraging code generation capabilities of LLMs}
  \label{fig:teaser}
\end{teaserfigure}

\maketitle

\section{Introduction}
\label{sec:intro}

Animation design lives at the intersection of visual design and code \cite{reynolds1982computer}. In application domains such as advertising, games, and user interfaces, creating animations requires a range of expertise in areas such as motion design, technical art, and front-end engineering. Like other creative work, animation design is inherently iterative, involving the concurrent specification and refinement of properties like movement and timing for multiple elements in a scene, which necessitates continuous adjustment and translation between artistic concepts and technical implementation. As a result, animation design, and the engineering needed to implement animations into applications, remains a challenging task for creators of all skill levels.

Given promising early results applying large language models to support designers in a range of artistic domains (such as visual design \cite{chiou2023designing, sanchez2023examining, choi2023creativeconnect}, creative writing \cite{gero2023social, suh2024luminate}, and 3D modeling \cite{gmeiner2023exploring, liu20233dall}), along with the ability of LLMs to quickly generate code to support software development\footnote{\url{https://github.com/features/copilot}}, we considered the potential of LLMs to support the animation design and engineering process. The ability of LLMs to interpret and respond to natural language requests may reduce the technical barriers needed to create animations, while also offering opportunities to support novices with both design and programming education \cite{jonsson2022cracking}. Yet, while use of natural language-based systems have proliferated in recent years (such as for \textit{text-to-image} generation with tools like Dall·E\footnote{\url{https://openai.com/dall-e-3}} and Midjourney\footnote{\url{https://www.midjourney.com/}} or \textit{code generation} with GitHub Copilot\footnote{\url{https://github.com/features/copilot}}), these existing commercial tools are not designed for the specific needs of animation creation because they lack support for the highly-iterative process of continuous refinement and visualization of animation designs. During this process, small changes to animation properties require a tight feedback loop of design, preview, and editing which is not currently supported with AI tools that require time to regenerate for every iteration and new prompt.

Further, while prompting strategies for text-to-image generators have been well studied in recent work \cite{chiou2023designing, sanchez2023examining, liu2022design, kulkarni2023word}, the transferability of these strategies to animation design is uncertain, as specifying how multiple items should animate together can be especially difficult to describe and plan for upfront. Therefore, to support animation design, we need alternative approaches to existing LLM-based design and programming tools, as well as deeper research insights into how animation designers might communicate their ideas and iteratively develop their work with the help of LLMs.
 
To explore design opportunities for LLM-supported animation design, we created an AI-powered animation tool called \system. With \system, users can create animated illustrations from static 2D images using a combination of prompting and direct editing of animation properties. Once a user provides an input scalable vector graphic (SVG) along with a natural language prompt, \system uses GPT-4\footnote{\url{https://openai.com/research/gpt-4}} to generate CSS animation code to animate the image. Because the generated animation is created using code, users can then directly edit animation properties with provided editors available in \system. They can continually iterate on their designs through additional prompts or direct edits, and they can also request variants from the LLM to ideate on new design directions. Through these features, \system accommodates users exploring and adapting their design goals as they iteratively construct an animation through combined prompting and editing actions, supporting the tight feedback loop needed for animation.

To inform the development of \system, we first conducted formative interviews with 9 professional designers and engineers who work on animations. Through these interviews, we aimed to answer the research question,

\begin{itemize}
    \item \textbf{RQ1}: \rqone
\end{itemize}

We used insights from these interviews to shape the design of \system and evaluated the efficacy of \system for supporting animation design through an exploratory user study with 13 users with a range of animation and programming experience. During this study, participants used \system to animate two provided illustrations over the course of a 90 minute session, with agency to decide how to prompt the system and refine their designs with \system. Our evaluation of \system examined the following two research questions:

\begin{itemize}
    \item \textbf{RQ2}: \rqtwo
    \item \textbf{RQ3}: \rqthree
\end{itemize}

Through our analysis of user prompting strategies, we contribute a categorization of semantic prompting styles users employed to describe motion with natural language, showing how \system empowered both novices and experts to create animations while focusing on high level design goals. We also describe how \system enables users to predominantly build up their designs through sequential `decomposed' prompting, where users iteratively develop their ideas through incremental prompts that animate individual elements within a scene, as guided by generated output. 

In addition, through providing an interface that enables users to request animation designs through natural language prompting, provides direct previews of animations alongside generated code, and enables direct edits to generated output, we show how \system allows for design exploration and iteration throughout the 2D animation design process. Our work makes the following contributions:

\begin{enumerate}
    \item \system, an open-sourced\footnote{\url{https://github.com/apple/ml-keyframer}} LLM-powered application for designing animations from static images.
    \item Insights into how iterative design practices with 2D animation were supported by \system, drawn from an empirical exploratory study with 13 users 
\end{enumerate}

\section{Related Work}
\label{sec:related-work}
Our work draws from related efforts considering how generative AI might support and create new opportunities for design, along with prior work underscoring the value of iteration in design processes.

\subsection{Generative AI in Design}
The introduction of LLMs has facilitated an unprecedented rise in commercial and research efforts to explore their application to design fields. Muller et al. argue that generative AI introduces new challenges to the field of HCI ``due to the serendipitous and uncertain nature of the design space,'' with numerous open questions about how to design user experiences that can effectively facilitate creative work \cite{muller2022genaichi}. Interface paradigms of co-creating with AI are being proposed in domains such as graphic design \cite{chiou2023designing, sanchez2023examining, feng2023promptmagician, liu2022design}, software development \cite{barke2023grounded, vaithilingam2022expectation}, creative computing \cite{angert2023spellburst, jonsson2022cracking}, UI design \cite{petridis2023promptinfuser}, writing \cite{gero2023social, suh2024luminate}, 3D CAD \cite{liu20233dall, gmeiner2023exploring}, and music \cite{louie2020novice}.

One common challenge with using natural-language-based AI design tools is learning to develop effective prompting strategies to steer generated output. The typical user journey for using generative tools today begins with crafting the overall structure of a prompt, followed by user assessment of generated designs and continuous refinement of prompts for better results \cite{mahdavi2024ai, xie2023prompt}. Several user prompt taxonomies have been proposed in the domain of text-to-image generators \cite{oppenlaender2023taxonomy, sanchez2023examining}, with artists frequently using modifiers specifying artistic style (e.g., `Cubism') and quality (e.g., `award winning'), along with keywords to spur surprising output (what Oppenlaender refers to as `magic terms' \cite{oppenlaender2023taxonomy}). Similarly, Chiou et al distinguish between `operational' keywords that specify concrete reference terms and `conceptual' keywords using abstract modifiers that are more likely to lead to unexpected results \cite{chiou2023designing}. We posit that animations involve complex and unique user considerations compared to images, including timing and coordination, that needs to be studied to best understand effective prompting guidelines for creating animated content.

Early work exploring the creation of animated content from natural language prompts include Spellburst, which provides creative coding support for p5js\footnote{\url{https://p5js.org/}}\cite{angert2023spellburst} and video generation with tools like RunwayML\footnote{\url{https://runwayml.com/}} and OpenAI's Sora\footnote{\url{https://openai.com/index/sora/}}. Yet, research exploring generative AI for creating animations from \textit{existing} image assets has been underexplored; this process is most closely aligned with current professional practices, where animations are built from assets created by visual designers to ensure high quality and consistency. Emerging research on LLM-generated scalable vector graphics (SVGs) suggests a promising path \cite{miriam2024text2illustration} but has been only applied to static images. Our system was designed and studies were conducted in late 2023, when, to the best of our knowledge, it was the first tool to apply LLMs to SVG-based animations. Since then, projects such as LogoMotion \cite{liu2025logomotion} and MoVer \cite{ma2025mover} have considered predominantly automated approaches for creating animated logos and text. However, recent work has demonstrated how the use of LLMs more broadly might contribute to homogenization of creative output or a reduction in critical thinking \cite{lee2025impact}, especially when LLMs present `complete-seeming' ideas \cite{anderson2024homogenization, bommasani2022picking}, which we argue automated approaches to animation lean towards. Thus, by focusing instead on how creators can specify animation goals from scratch (rather than in response to automated suggestions) our work points a way towards offering users more flexibility in crafting their creative vision.

\subsection{Iteration in Design Prototyping}
Iteration is a fundamental and crucial aspect of design process. Typically consisting of repeatedly exploring new ideas and refining designs, iteration can foster alternative perspectives \cite{dow2011prototyping}, reflection \cite{hartmann2006reflective, adams2003educating}, and stakeholder feedback \cite{dow2011prototyping, yen2017listen}, aiding designers in identifying challenges and uncovering new directions \cite{buxton1980iteration, kelly2002art, adams2002understanding, camburn2017design, dunlap2014heuristics}. Iteration is essential to complex designs where exploring different alternatives can reveal potential issues and opportunities. Animation design exemplifies a domain that greatly benefits from iteration due to its multiple stages and stakeholders. For example, film animations involves the integration of story, script, storyboarding, media design, and visual narratives \cite{tsai2013storyboard}, while animations for games can involve character rigging, lighting, and special effects \cite{heller2011becoming}. Creating animations often requires collaboration among diverse stakeholders and iteration to refine ideas and ensure alignment across all elements of the project. Our work with \system focuses on computer-based animation, specifically the creation of web-based animations, which often involves graphic designers and front-end engineers working together to create production-ready animations such as animated transitions, loading animations, and data visualizations \cite{drasner2017svg, head2016designing}.

Iteration involves both \textit{exploration} and \textit{refinement}, as illustrated by the classic ``Double Diamond'' model in design \cite{DesignCouncil}. Exploration typically entails generating ideas and multiple variants of a design solution, where designers experiment with different styles and design languages. Instead of committing to a single design concept at the early stage of the design process, creating design variants can help designers avoid the pitfall of fixation \cite{jansson1991design}, discover potentially valuable directions \cite{cross2004expertise, cross2006designerly}, and generate a more diverse array of ideas \cite{buxton2010sketching}. An effective practice of creating design variants is \textit{parallel prototyping}, which involves creating multiple design variants in parallel before making further improvements on any of the variants \cite{camburn2017design}. By generating and comparing varied prototypes in parallel, designers were able to create higher quality and more diverse design solutions and feel more confident in the design process than working on single designs \cite{dow2010parallel, dow2009effect, neeley2013building}. Refinement is where designers go deep on improving a single design to match their design goals \cite{ball1997problem}. It has been demonstrated in multiple studies that rapid refinement on existing designs can enhance design outcomes \cite{dow2009efficacy}. 
 
Despite the importance of iteration in design, commercial image generators like Dall-E \footnote{\url{https://openai.com/dall-e-3}} and Midjourney\footnote{\url{https://www.midjourney.com/}} are limited in supporting users with meaningfully iterating on their ideas. These tools often present users with polished visual output, which may lead users to narrow in on one design direction before fully exploring alternatives \cite{suh2024luminate}. Emerging research has started to explore ways to support iteration in design using generative AIs. For example, \cite{zhou2024understanding} presents a rule-based AI chat system for user interface design that facilitates the non-linear design process and generates alternatives for users to remix and refine. Presenting users with multiple options has been found to support users overcoming creative blocks in their process \cite{choi2023creativeconnect, angert2023spellburst}, spur new ideas \cite{tholander2023design, suh2024luminate}, and verify the quality of generated output \cite{barke2023grounded}. In the context of code generation, other related work has similarly distinguished between two use cases for LLMs: supporting `exploration,' when a user is not sure of what they want to create yet, and `acceleration,' where a task is well defined and the user wants the LLM to help them get to a solution faster \cite{barke2023grounded, angert2023spellburst, jonsson2022cracking}. While iteration and exploration are also crucial in animation designs, less is known about how generative AI can assist exploration and refinement in this domain.

In summary, research is needed to determine whether animations may require alternative prompting approaches compared to text-to-image generation strategies, or even to determine how effective LLMs may be at generating animations altogether. Our work examines how natural language prompting and code generation-based approaches could support animation design. Further, we consider how the integration of previews of LLM-generated designs, design variants, and code editing mechanisms might support iteration for animation design.
\section{Formative Study}
\label{sec:formative-study}

We conducted a formative interview study to answer our research question
\begin{itemize}
    \item \textbf{RQ1}: \rqone
\end{itemize}

 We recruited professional animation designers, developers, and prototypers via Slack channels dedicated to animation and front-end engineering at a large technology company\footnote{Both of our formative study (Section \ref{sec:formative-study}) and user study (Section \ref{sec:methodology}) were reviewed by the human subject research review board internal to our company. Selected participants received a \$12 meal voucher for per 45-minute they spent in our studies.}. 

During our 45-minute semi-structured interviews over video conference, we asked participants to share about their current animation design process and perspective of the role of generative AI for design. Finally, we probed the participants with a brief demo of an early prototype of \system to probe for feedback about LLM-based animation tools, with the facilitator demonstrating the system's ability to generate animations for a simple SVG using natural language (participants did not directly interact with \system). Participants received a \$12 meal voucher for their time.

Two researchers reviewed the session transcripts and applied thematic analysis \cite{braunUsing2006} to inductively identify a set of painpoints and opportunities across different stages of the animation creation process, as well as opportunities for generative AI support. Our protocol was reviewed and approved by the research ethics committee at the large technology company.

\subsection{Formative Interview Results}
We interviewed 9 participants (2 female, 7 male) representing a range of job titles such as designer, technical R\&D artist, and front-end developer. Their animation work includes user interface animations, advertising, instructional documentation, character animation, and interactive data visualization with between 3--22 years of professional experience (details about our participants can be found in Appendix \ref{app:formative-participants}).\footnote{We use FP to refer to formative interview participants and EP to refer to evaluative study participants in Section \ref{sec:user-study-results}} While a full discussion of challenges identified throughout the entire animation lifecycle is out of scope for this paper, we highlight a few opportunities especially pertinent to the role generative AI might play in animation design tooling.

\textbf{Challenge: Translating design to engineering implementation}.
Transitioning from design to engineering implementation (commonly referred to as ``handoff'') is time-consuming and can ultimately take away from other stages of the animation design process. Translation happens both because design is a collaborative process between designers and engineers, and because of differences in performance of animations across tools (e.g., production designs may appear differently than design mockups due to latency or rendering discrepancies).
Translation work, especially when it can involve several rounds of feedback and adjustment between prototyping and production, takes time away from design iteration: ``If I do one animation and try it and then it takes me 3 hours till I get to see it on the product, I don't get many cycles [of iteration] and then the animation quality drops significantly'' (FP3). 
Participants like FP1 imagined how generative tools that create both animation and the underlining code could mitigate this challenge: ``It would help because the code is there, so it will also help an engineer understand [the design]... it's a communication tool for both [engineering and design].''

\textbf{Opportunity: Supporting design iteration}. 
Animation creators saw potential in using AI to generate a starting point they could then refine: 
``If an [AI] could help create some more complex animations, or at least start it for me so I don't have to write every single line of code, that would save me so much time... I'm not writing the base code, I'm just truly putting that human touch to it'' (FP6) 
Animation creators also saw potential for generative AI to support iterating on animations, particularly by helping them explore multiple options. Participant FP7 imagined LLMs being most useful for refining designs rather than proposing completely new designs, suggesting using AI to apply design standards like accessible complementary colors or a dark mode from a light mode design. 
FP5 described how it was rare for animators to create a completely new timing curve, so applying animation properties from references examples would be valuable. Designer FP1 shared how AI can be especially helpful in generating design variants quickly from an initial design, tweaking designs so multiple could be compared at once. 

\textbf{Limitations: Lack of Creative Control}. A general concern about using generative AI was the risk of losing creative control, both in terms of reduced editing functionality along with a loss in creative satisfaction. For example, FP4 shared, ``[When] AI is being used to create the entire work, it doesn't scratch the itch that I think creative people have when they have that desire to create something. [With Midjourney] I feel like I'm pushing a button and something pops out.'' Participant FP9 also shared that, ``It's the granularity of customization that I really struggle with. When you're drawing, you control every stroke, and the nuance of it matters, and you control so little in AI by definition.'' Designers also feared having decreased understanding of how to edit generated designs: ``If you involve yourself deeply in all the details of the work, you have more freedom because you know exactly where things could break... [AI could] obfuscate some of these layers. That could make it harder, if something goes wrong, to know where to look because that code was essentially not written by you'' (FP8).

\subsection{Design Goals for AI-Supported Animation Tooling}
\label{sec:dgs}
Overall, our interviews revealed that animation creators were optimistic about using generative AI tools for quickly prototyping animation concepts, while still wanting ultimate control over the final design. Based on our findings from our formative interviews, we developed the following design goals (DG) for an AI-supported animation tool: 

\textbf{DG1: Support exploration for animations.} Animation creators imagined leveraging the power of AI to generate initial designs and to rapidly explore alternatives. Our system supports this goal by leveraging LLMs to generate initial starting points users can iterate on. We also take advantage of the flexibility of LLMs to allow users to generate multiple versions of a design at once to aid comparison and selection.   

\textbf{DG2: Enable granular controls for editing animations.} Our formative study revealed that animators perceive a lack of creative control in current text-based image generation AI systems. We also found that animation creators iteratively refine their designs when adapting to evolving design goals. As many animations are implemented in code, we anticipated opportunities to enable fine-grained refinement through direct or GUI-based edits to LLM-generated animation code.

\textbf{DG3: Empower non-experts to work with animation code.} We aimed to reduce the translation burden of turning designs mockups into implemented animations by leveraging the code generation power of LLMs, which we thought could be especially helpful for designers less familiar with code. Additionally, using natural language prompts to generate animations might open up opportunities for non-experts to get started with animation, which led us to expand our audience for testing \system to both professionals and beginners. 
\section{Keyframer System}
\label{sec:system}

\system is an LLM-powered application for creating animations from static images. Leveraging the code-generation capabilities of LLMs, along with the semantic structure of Static Vector Graphics (SVGs), Keyframer generates CSS for animating images based on user-supplied natural language prompts. Keyframer supports 1) \textbf{exploration} by enabling users to request and compare design variants and 2) \textbf{refinement}, as users can iterate on their designs either through reprompting or by editing the LLM-generated animations directly. Keyframer is implemented using GPT-4. 

\subsection{Animating an image with code generation}
With Keyframer, user provide an SVG image to animate, enter natural language prompts to generate CSS code, and can preview the resulting animations rendered from the generated code. 

\begin{figure*}
  \centering
  \includegraphics[width=0.8\linewidth]{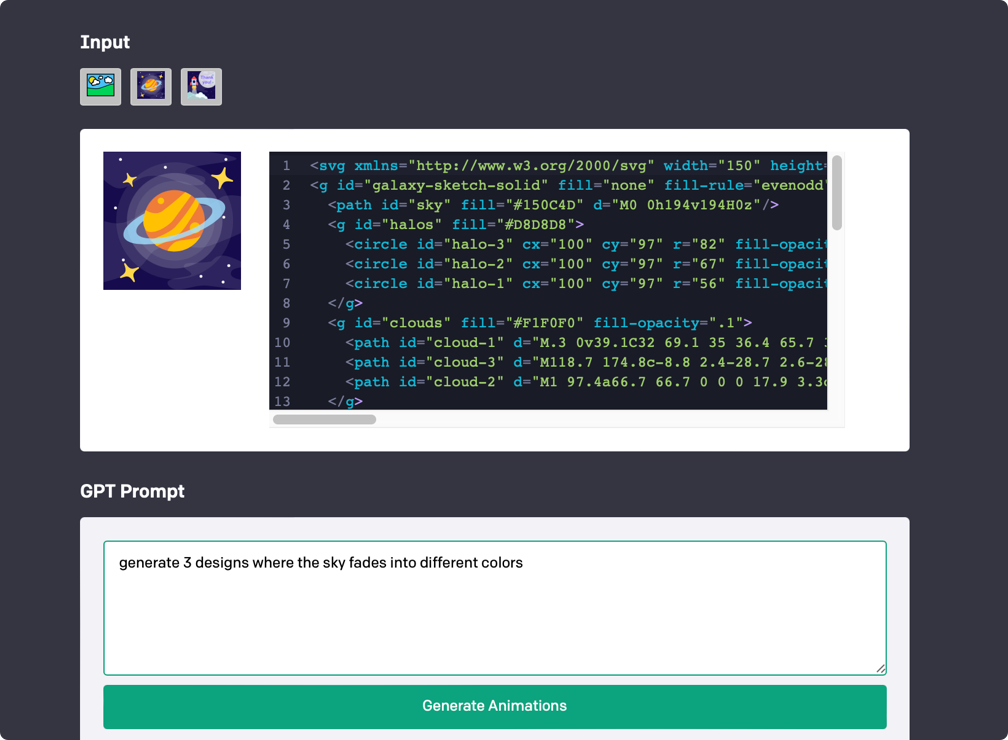}
  \caption{Image input field for adding SVG code and previewing image; GPT Prompt section for entering a natural language prompt.}
  \Description{SVG Input field for adding SVG code and previewing image; GPT Prompt section for entering natural language request for animation.}
  \label{fig:input}
\end{figure*}

The \textbf{Input} area is where users can paste the code for an SVG image they want to animate as shown in Figure \ref{fig:input}. SVG is a standard and popular image format used commonly in illustration and user interface animations\cite{drasner2017svg, head2016designing}. Because SVGs are XML-based images, the LLM is able to use descriptions of the scene embedded in the code, such as object identifiers. For example, in Figure \ref{fig:input}, the SVG code for the Saturn illustration contains identifiers such as sky, halos, and clouds. Object identifiers are commonly defined by designers by naming layers when creating graphics in applications like Illustrator or Sketch.

Following DG3, our system allows users to create animations by entering a natural language \textbf{user prompt} into a text input field. Users can make a request for a single design (``Make the planet spin'') or for multiple design variants (``Create 3 designs where the stars twinkle'', DG1). Before passing a user request to GPT, we supplement their prompt with the full SVG XML and specify the output format, instructing the LLM to return CSS animation code and ask for each code snippet to be accompanied by a descriptive explanation\footnote{The full details of our prompt can be found in Appendix \ref{app:append-full-prompt}}. We chose to use CSS rather than Javascript as CSS enables a wide range of animation effects and can be more interpretable for beginners (DG3). 

Once a request has been submitted, we stream the \textbf{GPT Output}, which consists of one or more CSS snippets as shown under ``\system Output'' in Figure \ref{fig:teaser}. Following DG3, our prompt asks the LLM to generates descriptive names for keyframes to make the code more interpretable for beginners (for example, adding CSS keyframe names like`flame-flicker' or `flash-sparkles'). 

\begin{figure*}
  \centering
  \includegraphics[width=0.8\linewidth]{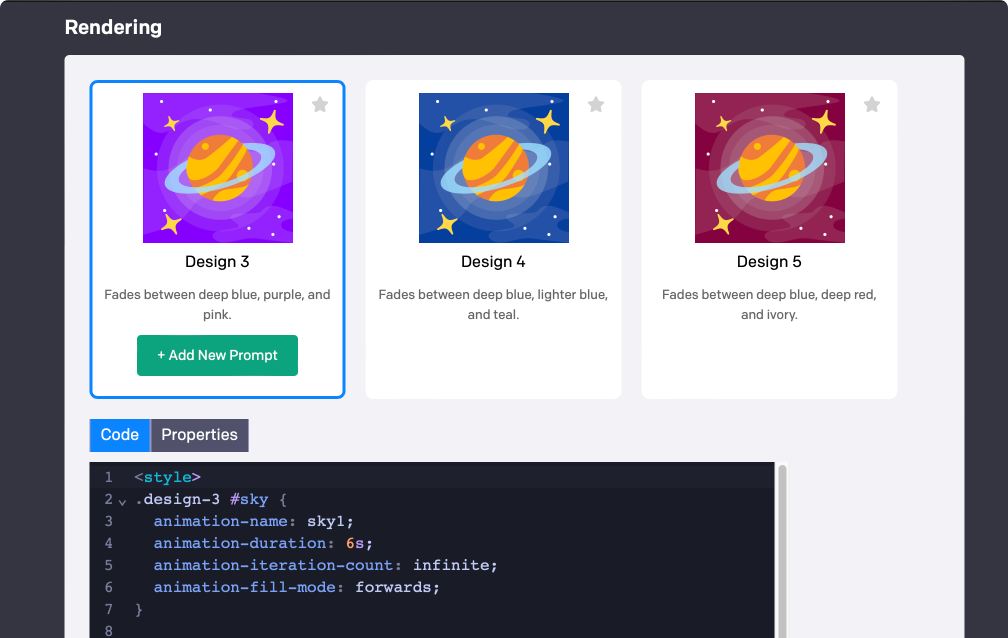}
  \caption{Rendering section for viewing generated designs side by side and editing output code in the Code or Properties editors.}
  \Description{Rendering section for viewing generated designs side by side and editing output code in Code or Properties editor.}
  \label{fig:rendering}
\end{figure*}

Each time we detect a complete snippet, we render a preview of the animation inline. Once a code snippet is detected, we inject the CSS snippet along with the original SVG code onto the page to display the animation. As we render multiple designs with the same underlying SVG on a single page, we have the LLM provide a unique class for each CSS snippet (e.g., \texttt{.design-1}, \texttt{.design-2}, etc.). When generating \textbf{multiple design variants}, visual renderings appear side by side to invite comparison and aid users in selecting the design that best matches their vision. For example, in Figure \ref{fig:rendering} where the user prompts ``generate 3 designs where the sky fades into different colors,'' the LLM generates three distinct options to spark ideas. 

\subsection{Editing generated animations}
Keyframer provides two editors for users to directly modify generated animation code (DG2):  a \textbf{Code Editor} where the CSS for a given design can be edited directly, and a \textbf{Properties Editor}, which presents dynamically created UI for editing CSS properties. Following DG3, the Properties Editor is designed to be approachable to users less familiar with code and is modeled after UI elements one might see in graphics editors like Illustrator.

\begin{figure*}
  \centering\includegraphics[width=\linewidth]{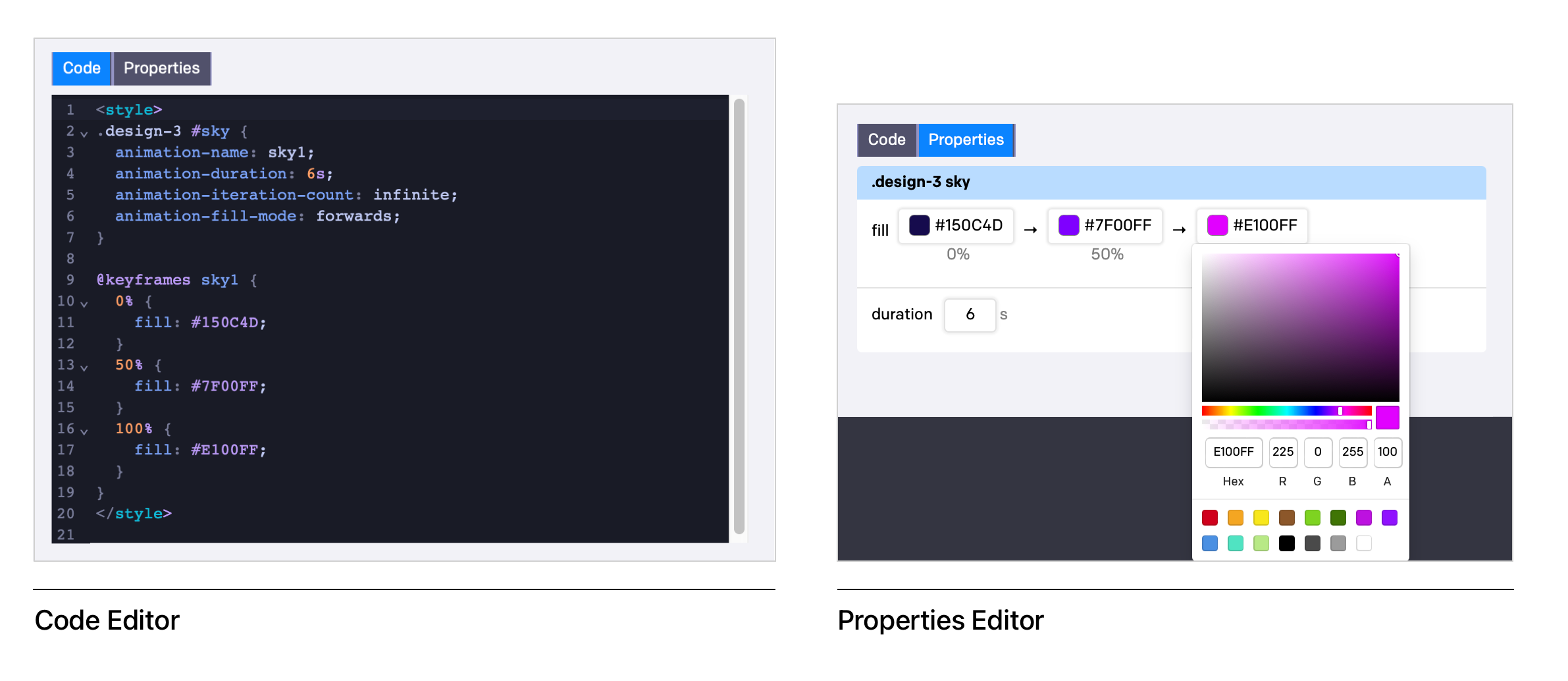}
  \caption{The Code and Properties Editors enable users to edit generated animation code directly. The editors are bi-directional so that any edits in one editor are reflected in the corresponding editor. The Properties editor adds a UI layer for editing animation properties without having to edit code.}
  \Description{The Code Editor and Properties Editors are bi-directional so that any edits in one editor are reflected in the corresponding editor.}
  \label{fig:editors}
\end{figure*}

The \textbf{Code Editor} is implemented with CodeMirror and has syntax highlighting and autocompletion for CSS. We re-inject the revised CSS onto the page on every keystroke so that users can preview their changes immediately. The \textbf{Properties Editor} provides property-specific UI for editing the code (DG3) such as a color picker and a dropdown and bezier editor for timing functions. Figure \ref{fig:editors} shows how an example code snippet appears in both the code editor and the properties editor. Edits in either mode are immediately reflected in the companion mode to support direct manipulation \cite{kery2020mage}.

\subsection{Iterating on animations}
\begin{figure*}
  \centering
\includegraphics[width=0.8\linewidth]{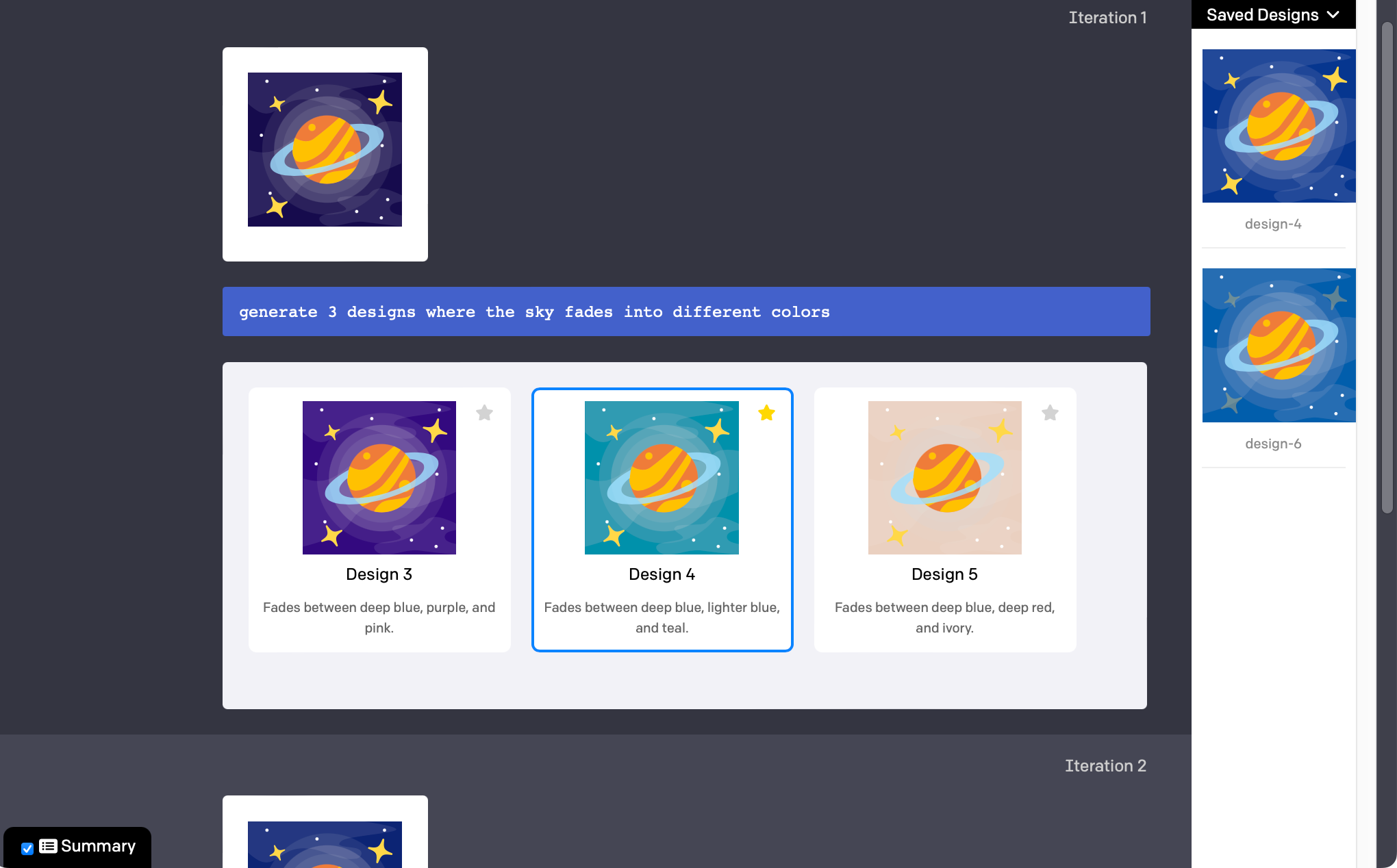}
  \caption{A summary view that hides all text editors and displays the user prompt and generated designs at each iteration. The Saved Designs sidebar (shown on the right) appears in either the Summary or regular view and shows all favorited designs. Clicking on a favorited design scrolls to the iteration in which the design was generated.}
  \Description{A summary view that hides all text editors and displays the user prompt and generated designs at each iteration}
  \label{fig:summary}
\end{figure*}

To support user exploration (DG1), \system allows users to iteratively build on a generated animation using additional prompts. Underneath every generated design is a \textbf{+ Add New Prompt} button; which lets users extend their design with a new prompt. In this way, users can continually add additional prompts in sequence to iterate on their animation. When a user adds a new iteration, our prompt requests the LLM to modify or extend existing CSS code as outlined in \ref{app:append-full-prompt}. To support review, users can bookmark their favorite animations or turn on a Summary mode to hide all text editors, allowing them to quickly revisit previous prompts and designs as shown in Figure \ref{fig:summary}.

\section{User Study Methodology}
\label{sec:methodology}

Our user study was designed to answer the following research questions:

\begin{itemize}
\item \textbf{RQ2} \rqtwo
\item \textbf{RQ3} \rqthree
\end{itemize}

\subsection{Study Procedures} 
We facilitated 90-minute sessions over video conference in which individuals built a set of animations with \system. Participants were presented with a hypothetical scenario in which they are asked to  help a children's book illustrator design animations for her personal website, using her illustrations. We began by briefly demoing \system, demonstrating how to enter prompts, request multiple designs, and use the editors to edit generated animations. We also introduced a third-party system called Boxy\footnote{\url{https://boxy-svg.com}} that could be used to explore the provided SVG images and read their element identifiers to use in prompting (see \ref{app:boxy} for further details).

\begin{figure*}
  \centering
  \includegraphics[width=0.6\linewidth]{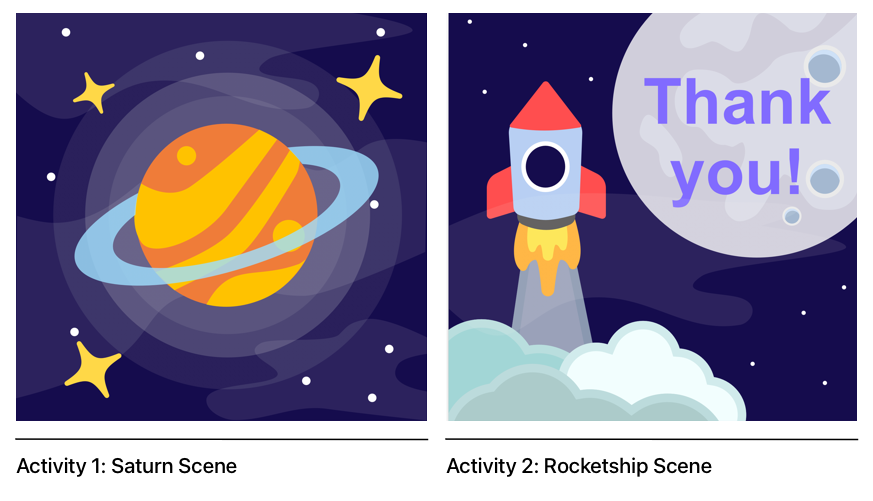}
  \caption{The two illustrations used for our user study activities. The Saturn illustration contains 20 unique elements, including a Saturn, its ring, and yellow `sparkles` (the three stars surrounding the planet). The Rocketship scene contains 19 elements, including a rocketship, clouds, and typography.}
  \Description{}
  \label{fig:illustrations}
\end{figure*}

We then started the main part of the study by asking participants to open Keyframer in their browser and create animations using two provided images (Saturn and Rocketship), displayed in Figure \ref{fig:illustrations}. Both SVG images were prepared by the authors of this paper in advance (full details on the creation of these SVGs can be found in \ref{app:illustration-creation}). For each activity, participants were first asked to explore the SVG using Boxy and brainstorm ideas for animating the image. Then the participants used \system freely for 15 minutes with the goal of creating designs they would be happy to share with the illustrator. We asked participants to think aloud as they crafted their prompts and responded to LLM output. All participants completed Activity 1 followed by Activity 2. 

After both activities, we used a semi-structured protocol to ask participants open-ended questions about their prompting strategies, feature requests, and potential use cases of using \system for future work. The full details of our study design, research instruments, and study format can be found in \ref{app:user-study}.

\subsection{Participant Recruitment}
We advertised our user study on a message channel with 2,500 members internal to a large technology company. In addition, we reached out via email to participants in our formative study for snowball sampling among their co-workers. 

Those interested in participating first filled out a screener survey, answering several questions about their job title, gender identity (for recruitment balancing purposes), any prior experience with animation or coding, and whether they had experience using AI tools. From the total pool of respondents, we selected a subset representing a range of programming and animation experience, as several professional animators interviewed in our formative study pointed out that \system might be especially helpful for novices.   
We mapped out a 2x2 matrix representing animation and programming experience such that we had representation from four groups: High Code and High Animation (HCHA), Low Code and Low Animation (LCLA), High Code Low Animation (HCLA), and Low Code High Animation (LCHA). A participant was considered low code if they had little or no prior programming experience; in contrast, a high code participant writes software professionally. Low animation users had little or no experience with animation design, while those categorized as high animation included users who had taken classes in animation design in college or do animation work professionally. 

\subsection{Data Collection \& Analysis}
We collected screen recordings and audio from each video conference session, giving us 19.5 hours of video for analysis. Additionally, we instrumented the application with an anonymized, session-based logging features to create timestamped activity logs tracking the prompts users entered, the LLM response for each request, and which editors participants used to edit the LLM output, if any. Transcripts auto-generated by the video conferencing software were reviewed and edited by the research team and after each session, researchers wrote reflective memos to capture salient moments. Our data collection and study protocol were reviewed and approved by a research ethics committee and legal counsel at a large technology company.

Following the procedures of thematic analysis \cite{braunUsing2006}, two researchers met weekly over the course of two months to analyze the transcripts and logs from our sessions to inductively identify themes around the \system user experience. We also quantitatively analyzed the prompts users wrote (which were coded blind of participant identifiers), the animations and code generated, and user editing patterns in modifying generated animation code, comparing across all four groups.  

We identified a set of prompting styles by first having two authors independently code a subset of 25 user prompts (or 12\% of all unique prompts) over multiple rounds. After reviewing and discussing any discrepancies in our application of these codes, the same two researchers split the list of prompts and independently coded half, discussing until we reached agreement on all prompts.

\section{User Study Results}
\label{sec:user-study-results}

In this section, we first share a summary of our study participants and their overall impressions of using \system. We then provide details about the performance of the LLM in generating animations, along with summary statistics about the designs and code generated. Next, we share results relating to prompting strategies (RQ2) and how \system supported iteration in animation design (RQ3).

\textbf{Study Participants} From 54 survey respondents, we selected 13 participants (6 female, 7 male), with 3 or 4 participants in each animation and code skill level category (Table \ref{tab:evaluative}) to gather qualitative feedback from a diverse range of perspectives. 
Those with `high animation' experience have worked on advertising campaigns, data visualization, and UI animations using After Effects (EP8, EP9, EP10, EP12, EP13) and Javascript and CSS (EP7, EP9, EP10). Four participants (EP1, EP2, EP3, EP13) had no prior programming experience. All reported having tried out existing AI tools (11 out of 13 have used ChatGPT, and 7 have used Dall·E and/or Midjourney), largely to test their capabilities and for fun.

\label{sec:results}
\begin{table}
  \caption{Evaluative User Study Participants}
  \label{tab:evaluative}
  \begin{tabular}{llll}
    \toprule
    Participant & Job Title & Group \\
    \midrule
    EP1 & Research Scientist & Low Code, Low Animation \\
    EP2 & Product Manager & Low Code, Low Animation \\
    EP3 & Brand Content Manager & Low Code, Low Animation \\
    \midrule
    EP4 & Software Engineering Manager & High Code, Low Animation \\
    EP5 & Machine Learning Engineer & High Code, Low Animation \\
    EP6 & Research Engineer & High Code, Low Animation \\
    \midrule 
    EP7 & Research Engineer & High Code, High Animation \\
    EP8 & AR/VR Engineer & High Code, High Animation \\
    EP9 & UX/UI Designer & High Code, High Animation \\
    EP10 & Creative Technologist & High Code, High Animation \\
    \midrule
    EP11 & Instructional Designer & Low Code, High Animation \\
    EP12 & Product Design Engineer & Low Code, High Animation \\
    EP13 & Motion Designer & Low Code, High Animation \\
  \bottomrule
\end{tabular}
\end{table}

\begin{figure*}[htb]
  \centering
\includegraphics[width=\linewidth]{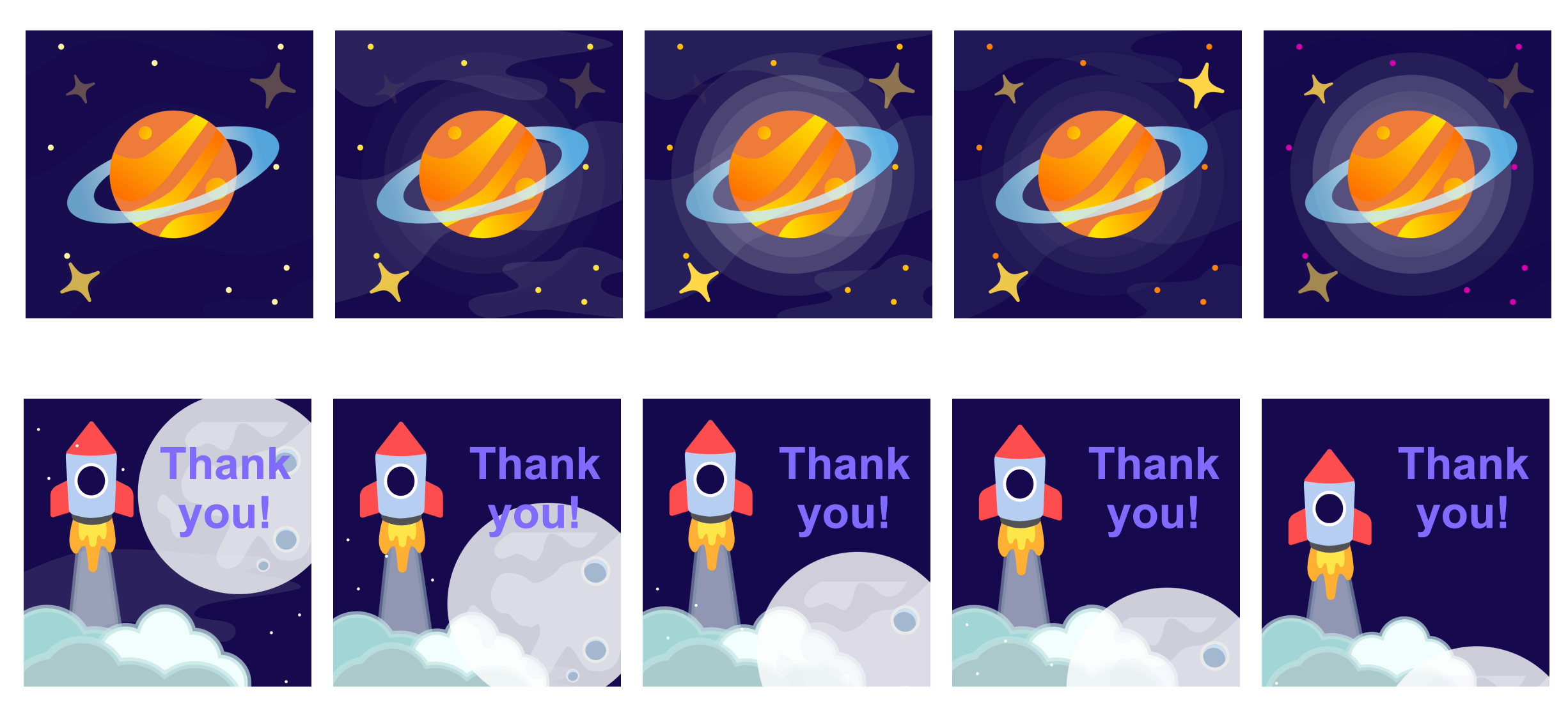}
  \caption{Five frames taken from animations generated by EP4 (HCLA) and EP9 (HCHA). The Activity 1 Saturn animation from EP4 has the sparkles fade in and out independently, the clouds fade in, the halos fade in one after another, and the specks alternate colors from yellow to orange to pink. The Activity 2 Rocketship animation from EP9 has the rocketship move up and down, the clouds grow and shrink in size, and items in the background (the moon and specks) shift down to give the appearance of the rocketship lifting off.}
  \Description{}
  \label{fig:results-examples}
\end{figure*}

\textbf{General Impressions of \system} Participants were happily surprised by the efficacy of the system to turn natural language instruction into animations. Beginners without animation experience found it easy to get started with \system: ``This is just so magical because I have no hope of doing such animations manually...I would find [it] very difficult to even know where to start with getting it to do this without this tool'' (EP6). Similarly, EP5 shared how \system ``democratize[d] this whole creative process'' and ``highlights creativity more so than the mechanics of [the code],'' letting users focus on higher level goals instead of having to know CSS syntax. Participants appreciated that the system sped up their animation prototyping process: ``Doing something like this before would have just taken hours'' (EP2).

Even professional motion designer EP13 saw the potential of \system to complement his animation process, sharing how LLMs are ``just another tool in our toolbox. It’s only going to improve our skills.'' 
 
\textbf{System Performance}. On average, $90.4$ (SD = $36.7$) lines of code were generated by the LLM in the final designs for Activity 1, and $87.2$ (SD = $63.5$) for Activity 2. The generated CSS from GPT-4 included CSS properties such as opacity, fill color, visibility, scale, and timing-function (a full list of generated CSS properties is listed in \ref{app:css-props}). \system generated animation code in a timely manner, taking on average $17.4$ seconds (SD = $9.7$s, max = $62.0$s, min = $5.9$s) to generate a code instance. In addition, the code generated by the LLM were generally high-quality, with only $6.7$\% (n=15) of code snippets with syntactical errors (the details of which are provided in \ref{app:css-errors}). An example of two participants' final animations are displayed in Figure \ref{fig:results-examples}.

\subsection{RQ2: \rqtwo}
\label{sec:results-rq2}

Each participant entered on average $15.8$ prompts (SD = $4.0$, min = 11, max = 26), with a mean word length of $16.7$ words per prompt (SD = $15.2$). The longest prompt consisted of $142$ words and the shortest contained only $2$ words (``remove exhaust''). 

In this section, we describe two different dimensions of prompting strategies we identified: 1) decomposed versus holistic prompting, and 2) high specificity vs semantic prompting.

\subsubsection{Decomposed vs Holistic Prompting}. We classified prompts as \textit{decomposed} if they involved animating elements one by one through sequential prompts. In contrast, \textit{holistic prompts} describe how multiple elements within a scene should be animated simultaneously within a single prompt. 

\begin{figure*}
  \centering
  \includegraphics[width=0.65\linewidth]{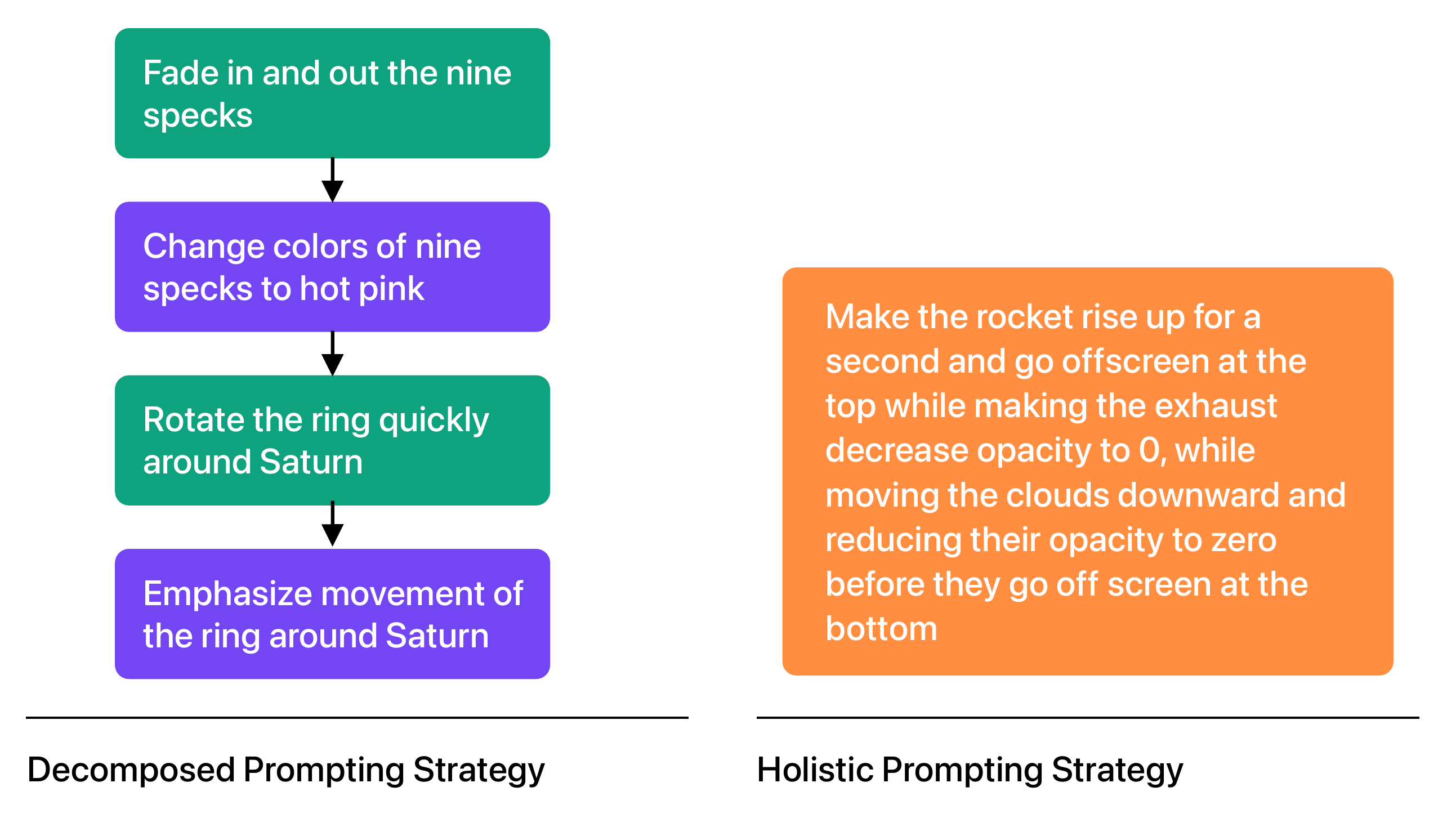}
  \caption{Decomposed prompts involve sequential prompting to animate and refine individual elements within an image, alternating between specifying animation for a new item in the scene and refining the generated animation. In contrast, holistic prompting specifies the entire behavior in a single prompt.}
  \Description{}
  \label{fig:decomposed-holistic}
\end{figure*}

All users took on a \textit{decomposed prompting} strategy for at least one activity, alternating between prompting for new animation behavior for an element and requesting refinements to the generated animation, as shown on the left of Figure \ref{fig:decomposed-holistic}. Through this process, users focused on a single feature and getting it to a state they were satisfied with before moving on to another element. This strategy enabled users to iteratively refine their animations through sequential prompting, rather than having to consider their entire design upfront in a single prompt. \system's ability to support decomposed prompting workflow is particularly effective for animation design as it enables designers to preview outputs, refine the movement of specific elements, and then proceed to the next step. Further, the generated output from an earlier prompt could seed future prompting; participant EP1, who is new to animations, shared, ``I'm reacting to what I see as its output and then adjusting on my end...it's informing and giving me ideas for what to explore next.'' Decomposed prompting enabled users to adaptively refine their design goals as they built up an animated scene through individual prompts.

In contrast, \textit{holistic prompting} was observed with 4 of the 13 users, with each user using this strategy during the second activity in particular\footnote{We believe holistic prompting may have been more popular in the second activity because the illustration lent itself to sequenced-based animation, e.g., specifying how items in the scene should respond to the rocket lifting off.}. With holistic prompting, users specified the behavior of multiple items simultaneously in a single prompt, as shown on the right in Figure \ref{fig:decomposed-holistic}. Software engineer EP4 described this strategy by comparing it to sculpting: ``It was probably easier to try and get the rough shape of it all at the very beginning [in a single prompt], and then fine tune the details versus trying to tell it [the LLM] cut by cut.'' The other 3 participants used holistic prompting to save time or because they felt it was necessary for having items move simultaneously. However, by revising the same underlying prompt, users run the risk of overriding desired behavior.

\subsubsection{High Specificity vs Semantic Prompting}. We observed two different styles of describing animations. \textit{High specificity prompts} ($34.6$\% of all prompts) are characterized by using animation keywords (like opacity, rotate, and scale) as well as expected values (e.g., `between 0-1.5 seconds'). In contrast, the majority of prompts (84.4\%) were \textit{semantic}, which were less defined and more descriptive (e.g., `make the clouds wiggle').\footnote{We qualitatively coded the ``high specificity'' and ``semantic'' features among the 205 unique prompts used across all users. These features were not mutually exclusive, as some prompts could contain high specificity for one element and semantic descriptors for another; thus the percentages sum up over 100\%.} We present a categorization of semantic prompt types in Table \ref{tab:semantic-prompts}, with categories 1) describing the movement, timing, direction, coordination, or other visual qualities of objects in the scene, and 2) the phrasing of those requests through incorporating analogies, using indefinite values, or asking for ideas (e.g., `make it look cool'). 

\begin{table}[htb]
  \caption{Categorization of Semantic Prompt Types}
  \label{tab:semantic-prompts}
  \begin{tabular}{l p{4.5cm} c p{5.25cm}}
    \toprule
    Prompt Type & Definition & Frequency & Examples \\
    \midrule
    Movement & Describing transformation (translation, rotation, scaling, etc.) & 68.4\% & twinkle, wobble, balloon upwards \\
    Timing & Describing speed and timing curves & 27.5\% & slowly, very fast, with an elastic feel \\ 
    Direction & Describing relative placement and orientation & 30.1\% & from below, vertically on the same plane \\
    Coordination & Relationship and synchronization between items & 19.7\% & together, keep attached, follow \\ 
    Other Visual & Opacity, color change, fonts & 8.8\% & increasingly transparent, alternate color \\
    Analogies & Similes or metaphors & 2.6\% & like stars in the sky, like an active fire \\
    Indefinite Values & Indeterminate values & 14.0\% & at different rates, random value \\
    Ideas & Requesting new ideas from the LLM & 11.9\% & do something crazy, show alternatives \\    
  \bottomrule
\end{tabular}
\end{table}

Participants were surprised by how well the LLM could interpret semantic prompts: ``I was generally impressed by how ambiguous I could be with my descriptions. It usually picked the correct or reasonable transform to use'' (EP7). Similarly, EP8, who used the prompt  \textit{Give me 3 designs where the clouds wiggle} shared: ``The fact that it got make the clouds wiggle in an interesting way that I was happy with is kind of wild.'' Being less specific opened up creative possibilities: ``If you could be more vague [in your prompts], then you have more options for surprise and delight'' (EP4). For example, EP5 utilized open-ended prompts (e.g.,``Make it look cool'') to curate new ideas: ``Instead of explaining the steps of making it look as cool as possible at an atomic level...you can just say what the goal is and [the LLM] can just fill in the rest.'' For others, semantic prompts saved time, reducing the need to define all parameters upfront. Participant EP8 transitioned from high specificity prompts in Activity 1 (e.g., \textit{For each of the specks, animate them pulsing between 85\% and 115\% of their current size. Each pulse should take between .5 and 1.5 seconds''}) to more semantic prompts by Activity 2 (e.g., \textit{Move the rocket exhaust and clouds from the bottom to the top}), sharing how generated properties like duration were often reasonable as is or were easy enough to directly edit, thus not requiring full specification upfront.

\subsubsection{Debugging Prompts}. While participants were frequently surprised that Keyframer understood their prompts, they expressed how articulating motion in natural language was challenging in and of itself. There were several moments where the LLM output did not match a user's expectations, yet they acknowledged that the LLM had a reasonable interpretation: ``It's definitely that sort of language barrier thing where it did exactly what I wanted, but it turns out that wasn't what I wanted. It did exactly what I told it to do, but I should have told it to do something else'' (EP4). Several users wanted the LLM to confirm its understanding before generating designs by asking the user to select one of multiple interpretations. Others proposed having the system make suggestions \textit{during} the prompting process by highlighting corresponding keywords from the prompt alongside its corresponding SVG element as they typed, indicating whether the LLM could correctly link words in their prompt to items in the scene.
 
\subsection{RQ3: \rqthree}

In the following sections, we focus on how \system shapes users' iterative process---including \textit{exploration} and \textit{refinement}---on their animation designs.

\subsubsection{Exploration through unexpected output and generating variants} 
\hfill\\
Participants found \system useful for helping them brainstorm directions for animations, both because of unexpected design it generated along with the use of the multiple variants feature. At times, \system surprised participants with a design they had never thought of, which spurred new ideas: ``Even if it doesn't come out the way you want it to, it's showing you another path that you can take... I think it's helpful to just come up with some other creative solutions'' (EP3). \system could also offer new directions for technical implementation. For instance, when EP7 prompted the LLM to animate the flame of the rocket as it launches, the system returned a vertically scaling effect that surprised him: ``I thought I’d have to morph the flame to get something like this effect...I was like, `Oh, this isn't bad either.' It's basically like another person suggesting something for me to try.'' 

A total of 9 out of the $13$ participants also explored ideas by explicitly ask the LLM to generate multiple designs. This design feature helped users brainstorm and compare ideas: ``It was nice to see the different possibilities and then feel like I could choose among them which one was more closely aligned'' (EP1). 

Generating multiple designs could effectively support different stages of the animation design process. When participants were initially exploring what to animate, they could leverage the multiples feature to map different directions: ``If I could get it [the LLM] to initially show me a larger set of possibilities, that would probably help me change the vision of what I want'' (EP4). In later stages, participants valued using variations to support detailed refinement like timing and duration of animated elements. 

At the same time, some participants expressed hesitancy about generating multiple designs in practice. When participants already had a clear idea of what they wanted, they would not generate multiple designs as it might steer them away from their vision. Some wished that the variants could better match their designs goals, be more visually distinct from one another, and take less time to generate, with EP2, EP5, and EP7 expressing that sometimes only 1 or 2 of the generated variants were relevant to their request.

\subsubsection{Refinement through combined prompting and direct editing}
\hfill\\
\system's editors supports participants with a range of programming experience with editing LLM-generated code. A majority of participants (69.2\%) edited the LLM-generated code with either of \system's editors. Seven participants edited code using the Code Editor, while $7$ made edits with the Properties Editor. On average, each participant edited $25.1$\% (SD = $0.2$) of the instances where they received generated code. Notably, 4 out of 6 Low Code users edited output code, and of the 5 High Code users who edited LLM output, the majority (80.0\%) used the Properties editor, suggesting that the Properties editor can be beneficial to both experts and novices. In fact, most participants that used the editors used \textit{both} the Code and Properties editor throughout their use of \system (55.6\%), suggesting that each editor can be beneficial regardless of prior experience with animation or code. 

Participants described how it was the combined powers of prompting and code editing that made \system helpful for refinement. A common strategy was to first use natural language to prompt for a design close to what a user had imagined and then leverage the editing interfaces to adjust details: ``I feel like the natural language part gets me 95\% of the way there, and then the last 5\% I would prefer to do edits here [the Code Editor]'' (EP9). Participants also shared that getting boilerplate code generated from natural language could leave them more time to focus and iterate on the design itself. Especially when having clearly defined goals, participants preferred to edit the generated code directly compared to re-prompting the system.
\section{Discussion}
\label{sec:discussion}

We found that a key aspect of \system's design was the combined features of supporting semantic natural language prompts, along with
direct editing of the rendered animations through code. Natural language input can support users in specifying their goals, and the majority of prompts users created (84.4\%) were semantic prompts that describe parameters like visual effects, timing, and coordination without using keywords commonly used in animation software. Simultaneously, even ``low code'' users found that direct modification of the LLM-generated code and parameters was helpful for fine-tuning their designs. This means that in the context of using LLMs to generate code-based visual artifacts, tool developers should spend effort building direct manipulation interfaces and not simply fall back to a simpler chat-only design.

Our participants appreciated the ability to progressively construct their prompts, generating their animated creations over multiple iterations. This contrasts with the one-shot interfaces commonly found in text-to-image generators, where users must describe their goals in a detailed, single prompt \cite{kulkarni2023word} and where image quality is often correlated with lengthy prompts \cite{xie2023prompt}. This was particularly enabled by the choice to save the generated artifacts at each prompt iteration, and to allow users to go back and create a new prompt starting with any previous iteration. Not only did this implementation choice support progressive iteration, but it also enabled the ``decomposed'' prompting strategy that was used by the majority of our participants. This ability to construct from scratch and adapt design goals over time is in contrast to related animation work that centers on automated approaches to generating animations \cite{liu2025logomotion, ma2025mover}, which may run the risk of design fixation in response to seemingly-fully-realized designs \cite{anderson2024homogenization, suh2024luminate}.

Explicit support for multiple variations also assists in the iterative process and is another feature of \system that should be included where possible in future animation tools. We found this feature was most useful in the exploratory design phase, but could be a hindrance during fine-tuning later in the design process. Much work remains to build even better tools for generating and displaying multiple variations. A particularly fruitful area for future work might be to explore methods for determining the best number of variations to generate based on where the user is in the design process, ensuring that those variations are clearly distinct, and finding better ways to visualize differences between variations when they are subtle.

There remain several open questions about how LLMs might be effectively integrated into animation design processes.

\textbf{Enabling user feedback on model interpretability} In the context of applying our categorization of semantic prompts, we imagine opportunities for systems to provide enhanced user feedback on model interpretability of users prompts. For example, \textit{during prompting}, a system might proactively display its understanding by both highlighting keywords in the user prompt along with elements in the illustration, which may help users confirm how LLMs are matching their prompt keywords with individual components within a graphic. 
\textit{Before generating a design}, some users in our study suggested having the LLM ask questions to confirm its understanding, which may assist users with clarifying their intention through revising their wording. \textit{After generation}, features like Keyframer's ability to display variants could be used to expose the range of ways an LLM might interpret their prompt, enabling users to choose from several options the one that best fits their vision. 

\textbf{Supporting task decomposition for non-linear design processes}  While Keyframer was designed to resemble the ChatGPT interface in which prompting happens sequentially and linearly, this linear representation was not necessarily representative of how some of our users approach animations. For example, EP12 stated that he would typically tune the movement of two items independently and simultaneously throughout the process of creating an animation. Future work integrating non-linear representation may help users expecting to tweak and merge different animated elements together, as explored in related work \cite{angert2023spellburst, dang2023worldsmith}. Interfaces that visually represent task decomposition on individual elements within a scene may enable more granular and flexible refinement.

\textbf{Enabling edits on underlying graphics} A common user request was to edit the underlying illustration using the LLM, which is not easily achieved with CSS alone. Not being able to edit the SVG itself prevents actions like adding new items to a scene, modifying the shape of objects, or re-grouping objects. While an earlier version of Keyframer enabled generating CSS \textit{and} SVG edits, it led to much longer response times and likelihood of exceeding token limitations because of the amount of code in the SVG itself. Future work can consider how LLMs can be used to edit both animation properties and the illustration itself to offer even more creative opportunities.

\textbf{Mixing prompting and direct manipulation} Many animators are used to direct manipulation via timeline-based interfaces such as those available in After Effects. For example, EP13 described how transform operations can be faster through direct manipulation, while more complex, non-translational animations might be easier through prompting. An open question is how to design future AI-powered animation tools combining the affordances of direct manipulation (such as that offered with \cite{masson2024directgpt}) with prompting for streamlining the definition and editing of animated properties.


\textbf{Designing animations with interactivity} Finally, in \system we explored the creation of animations for non-interactive contexts where the animation always plays exactly the same way every time. Other animation tasks, such as designing how a button appears when pressed or how a user interface responds to a gesture, need to take the user's interaction into account during playback. This type of task is more complicated, as it requires anticipating possible interactions and viewing previews of diverse interactions that affect the animation design in different ways. LLMs might serve an additional purpose in such a tool by helping both with design as in \system, but also in helping the designer understand possible interactions and identify edge cases in their designs.
\section{Limitations and Future Work}
\label{sec:limitations}
There are several limitations in what \system supports that could be expanded in future work. First, a common user request was to be able to edit the underlying illustration, which is not easily achieved with CSS alone. Not being able to edit the SVG prevents actions like adding new items to a scene or modifying the shape of objects, which may limit a user's creativity. However, while an earlier version of Keyframer enabled generating CSS \textit{and} SVG edits, it led to much longer response times and likelihood of exceeding token limitations because of the amount of code in the SVG itself. So while it might be ideal for users to leverage the LLM to make changes to illustration metadata directly, doing so can impact system performance.

Combining direct manipulation with prompting, such as described in \cite{masson2024directgpt} and \citet{huang2024plantography}, may streamline the definition and editing of animated properties. For example, EP13 described how transforms may be faster through direct manipulation, while more complex animations might be easier through prompting. Further, future work could translate text-based CSS animation code to a visual timelines that may better coordination of timing between elements, such as that used in Chrome's Animations panel UI.\footnote{\url{https://developer.chrome.com/docs/devtools/css/animations/\#animations-ui}} 

We utilized GPT-4 as our underlying LLM because of its availability and performance at the time of our experiments, but new models may yield different prompting approaches. In particular, multimodal models are now generally available, but were not at the time of this work. We used SVG as our image format because of its text-driven nature and its feature of labeling layers with natural language strings, which worked well with the text-only LLM that we used in \system. Newer systems built on multimodal models could have far greater flexibility. 

Finally, further work is needed to understand prompting practices for more complex illustrations, such as those used in UI animations and transitions. Yet, we believe that animated 2D illustrations such as those used in our user studies are representative of many potential applications like animated album art, emojis, or user avatars. 
\section{Conclusion}
\label{sec:conclusion}

Through \system, we envisioned how future animation design tools could be designed to support highly semantic natural language prompting, leverage and preview the outcomes of code generation, and enable design iteration through direct editing. 

Through our user study, we contribute a categorization of semantic prompting styles observed from users describing motion in natural language. We revealed ways to provide pathways for iteration, where users can alternate between prompting and editing generated animation code to refine their designs. We also unpacked how animation creators can gradually build up and explore ideas through sequential `decomposed' prompting. Through this work, we hope to inspire future animation tools that leverage the powerful generative capabilities of LLMs to expedite design prototyping, with dynamic editors that enable creators to maintain creative control throughout their design process.

\bibliographystyle{ACM-Reference-Format}
\bibliography{references}

\appendix

\section{Appendix}

\subsection{Formative Interviews Participants}
\label{app:formative-participants}

Our formative interview participants consisted of professionals who create animators as part of their work at a large technology company, as displayed in Table \ref{tab:formative}.

\begin{table}[htb]
  \caption{Formative User Study Participants}
  \label{tab:formative}
  \begin{tabular}{lllcl}
    \toprule
    Participant & Job Title & Animation Domain & Years of Experience & Animation Tools \\
    \midrule
    FP1 & Designer & UI Animation & 8 & After Effects, Cinema 4D\\
    FP2 & Creative Technologist & Advertising & 22 & Javascript, After Effects\\
    FP3 & Technical R\&D Artist & Character Design & 10 & Maya, Blender\\
    FP4 & Front-end Developer & Advertising & 7 & Javascript, After Effects\\
    FP5 & Software Engineer & Instructional Content & 6 & After Effects\\
    FP6 & UI Prototype Engineer & UI Animation & 15 & Swift, After Effects\\
    FP7 & Research Engineer & Data Viz & 8 & JavaScript\\
    FP8 & Research Engineer & Data Viz & 3 & Swift\\
    FP9 & Web Technologist & Data Viz \& UI Animations & 20 & Javascript, Unity \\
  \bottomrule
\end{tabular}
\end{table}

\subsection{Illustration Creation}
\label{app:illustration-creation}


We created two illustrations for users to animate, shown in Figure \ref{fig:illustrations}. We intentionally designed the two illustrations to help us observe a range of user prompting styles. For the Saturn illustration, multiple items can be animated simultaneously without timing dependencies. But for the Rocketship illustration, users might want items to move elements together in sequence (such as the rocketship and clouds moving together, followed by the text popping up). Given that we anticipated that prompting for timing between elements may be more difficult, we had users first animate the Saturn scene followed by the Rocketship illustration.


\textbf{SVG Pre-Processing}. For our user study, we did not expect users to need to have any knowledge of SVG syntax to successfully complete our tasks --- we wanted users to pay attention only  to the identifiers for items in the scene (which could be used for prompting the LLM). All other SVG properties (like fills or path origins) could be ignored. Additionally, to reduce the LLM response time and likelihood of encountering token limitations (as the SVG itself is passed in each request to GPT-4), we took efforts to make the SVGs as small as possible by reducing the lines of code in the XML. To accomplish both of these goals, we pre-processed the SVGs through the following steps:

\textbf{Step 1: Create Illustrations in Sketch}. We created the original illustrations in the design software Sketch. In Sketch, we gave layers descriptive names (e.g., `saturn' or `halos') and grouped them as appropriate (e.g., a `halos` group included `halo-', `halo-2', and `halo-3'). We then exported each illustration as an SVG.

\textbf{Step 2: Remove Inline Transforms}. We used Inkscape (https://inkscape.org/), an open-source vector graphics tool, along with the Inkscape extension Apply Transforms (https://github.com/Klowner/inkscape-applytransforms) to remove transform definitions within the SVG XML. This step was taken to minimize potential issues with applying CSS transforms like rotations that could conflict with transform properties already defined in the SVG.

\textbf{Step 3: Minimize SVG}. We used the tool SVGOMG (https://jakearchibald.github.io/svgomg/) to simplify the SVG by removing unnecessary metadata like docstyle or XML instructions not needed for our study.

\textbf{Step 4: Reformatted SVG Identifiers}. We manually edited the SVG code to move all identifiers to the beginning of an element definition (e.g., \texttt{<path id="sky".../>}) so that when users are quickly inspecting the SVG code, they can easily find the id of each element without having to scroll horizontally in the code editor.

After pre-processing, the SVG for Activity 1 consisted of 42 lines of code, and the SVG for Activity 2 consisted of 40 lines of code, translating to 2,503 and 3,667 tokens respectively for GPT-4\footnote{https://platform.openai.com/tokenizer}.

\subsection{Using Boxy for Reviewing SVG Elements}
\label{app:boxy}

\begin{figure*}
  \centering
  \includegraphics[width=\linewidth]{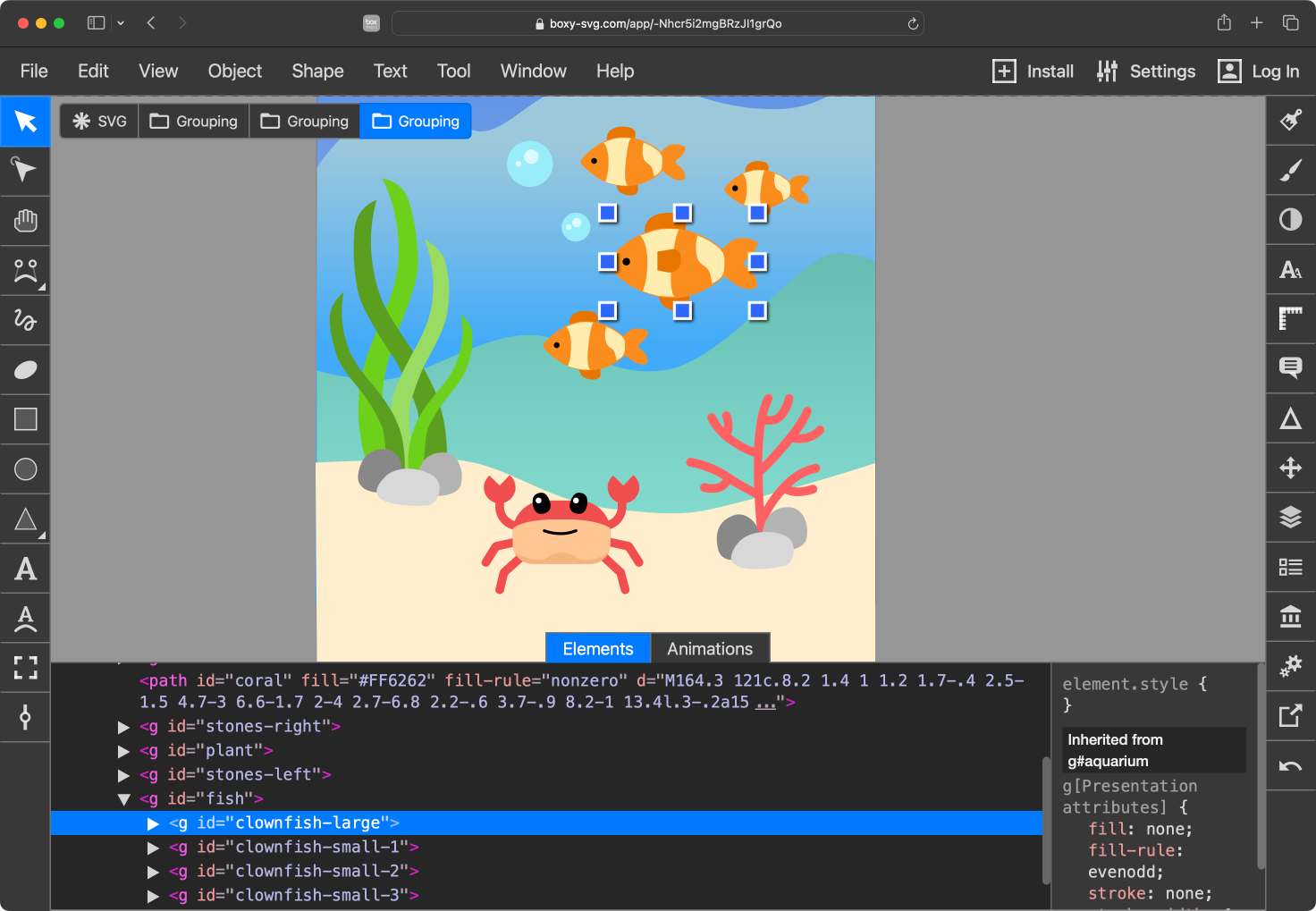}
  \caption{Boxy, a 3rd-party application we used for participants to explore each SVG illustration before using Keyframer. Clicking on an item in the scene reveals the corresponding line of code and its identifier (in this example `clownfish-large' for the selected fish). Likewise, clicking on a line of code will highlight the corresponding item in the illustration.}
  \label{fig:boxy}
\end{figure*}

We used a web-based third party tool called Boxy\footnote{https://boxy-svg.com/} for participants to interactively inspect the SVG so they could familiarize themselves with the identifiers of items in the illustration. These identifiers could then be used in Keyframer to animate the scene via natural language prompts. We loaded each illustration into Boxy, which can display the SVG illustration alongside its code as shown in Figure \ref{fig:boxy}. With Boxy, users can click on items within the image to reveal its identifier. You can also click on a line of code to highlight the corresponding item in the illustration.

\subsection{User Study Procedures}
\label{app:user-study}
Each facilitated 90-minute session involved the following protocol:

\textbf{Intros} (15 min) Brief introductions and several questions about the participant's background (including what they do for work and what prior experience they may have with animation and programming)

\textbf{Boxy Intro} (3 min) The facilitator shared their screen and gave a brief demo of Boxy with a sample illustration distinct from the two used in the user study task (Figure \ref{fig:boxy}). They demonstrated both how to click on items in the image to see the corresponding code, and how to click on lines of code to highlight the item in the illustration.

\textbf{Keyframer Demo} (5 min) The facilitator walked through the Keyframer interface and described how to use it through an example. The example used an SVG whose scene consists of a sun, several clouds, a sky, and a hill. The facilitator first entered the prompt `\texttt{animate the sun to move from the top left to the bottom right}.' Once the LLM fulfilled this request, they showed how you can either edit the output animation through the code editor or the property editor. Next, the facilitator demonstrated how you can continue animating the scene by creating a new iteration. They then entered the prompt `\texttt{generate 3 designs where the sky changes colors like a sunset}' to show how you can request multiple examples from the LLM. Once three designs variants appeared, the facilitator showed how custom UI elements appear for different CSS properties (in this case, showing how you can edit color properties generated by the LLM using a color picker). The facilitator then asked the participant if they have any questions before proceeding.

\textbf{Activity 1 Boxy Exploration} (5 min). The facilitator introduced the design scenario and sent the participant a link to Boxy to inspect the Saturn illustration (Figure \ref{fig:illustrations}). At this point, the participant shared their screen for the remainder of the session. The participant was asked to take 5 minutes to explore the SVG in Boxy to get the name of elements in the scene and come up with a few ideas for what they might want to animate, thinking aloud as they worked.

\textbf{Activity 1 Keyframer Exploration} (15 min). The facilitator sent the participant a link to the Keyframer app they could open in their web browser. They had 15 minutes to come up with a few animation designs they are happy with and could share with their friend. During this time, the participant entered as many prompts as they wanted to create their design.

\textbf{Activity 1 Interview} (5-10 min) The facilitator asked the participant several questions about their experience including: What went well? What did not go as expected? Did anything surprise you about using the LLM?

\textbf{Activity 2 Boxy Exploration, Keyframer Exploration, and Interview} (5 min, 15 min, 5-10 min) We repeat the same process for the second illustration (Rocketship scene) where the participant was asked to create several animations they are happy with, followed by the same questions about their experience.

\textbf{Closing Interview} (10 min) We closed the session asking several questions about their overall experience.  First, we asked users to rate on a scale from 1-5 their response to three questions: 1) How satisfied are you with how far you were able to get with your animation in 15 minutes?, 2) How helpful do you think this system was for helping you brainstorm different ideas?, and 3) How helpful do you think this system was for helping you refine your animations? (with 1 being very unsatisfied, 3 being neutral, and 5 being very satisfied). We then followed with several open-ended questions, including: Do you think you discovered any strategies for prompting that seemed to work better? Are there features you would want this tool be able to do? Could you imagine using this type of technology for animation design in the future? All participants received two \$12 cafeteria vouchers for their time.

\subsection{Full Prompt for Generating CSS Animation Code with an Input SVG}
\label{app:append-full-prompt}

\begin{shaded}
\begin{lstlisting}[frame=none]
 You are writing CSS for animating the SVG contained within the <> symbols below. The design should meet the following user specification:
    
    ###
    <User-entered prompt>
    ##

    Please follow these rules in writing the CSS:
    1. Contain the CSS code snippet in a <style> element.
    2. In the CSS code snippet, do not use animation shorthand and use the property "animation-name".
    3. If there is any transform: rotate() or transform: scale() in the code snippet, set transform-origin: center and transform-box: fill-box.
    4. The animation should repeat forever with animation-iteration-count: infinite.
    5. The CSS for a code snippet should be nested within a parent with the class design-n where n corresponds to the index of the snippet counting up from <number of existing designs>. ONlY add this parent-class to CSS rules. DO NOT add the parent class to keyframes.
    6. A code snippet should be followed by a short explanation summarizing how the design is distinct. The explanation should be no more than 15 words long, should be descriptive rather than technical, and should be contained in an <explanation> tag.
    7. Only write CSS. Do not return any SVG or additional text.

    If the user asks for more than one design, follow these addition rules FOR EACH DESIGN: 
    1. Generate independent CSS code snippets and explanations for each design.  
    2. End each CSS code snippet and explanation with the delimiter -----. Ensure the delimeter has 5 dashes.

    <>
     <The SVG Code>
    <>

\end{lstlisting}
\end{shaded}

If the user is working in a new iteration (e.g., the generated code should build off of a selected design), we also include these instructions for extending the existing CSS code:

\begin{shaded}
\begin{lstlisting}[frame=none]

    For all generated designs, start from the existing add any new CSS BELOW the existing CSS.
    
    """<Existing CSS>"""
    
    Refactor the existing CSS to apply the corresponding design-<number-of-existing-designs> class.
    RETAIN ALL LINE OF IN THE EXISTING CSS. DO NOT DROP ANY LINES.
    Check your work and ensure that the existing CSS is represented in the designs.
 
\end{lstlisting}
\end{shaded}

Note that while the prompt asks the LLM to retain existing CSS and only append new CSS below, in practice we found that the LLM often modifies existing code when it is more efficient. For example, if a user prompts the system to increase the speed for a particular animated feature, the LLM updates the `animation-duration' property of the existing CSS code, rather than adding an override for the timing below it. 

\subsection{Generated CSS Properties}
\label{app:css-props}
Table \ref{tab:css-props} presents the range of GPT-4 generated CSS property types in its responses to user prompts:

\begin{table}[htb]
  \caption{Generated CSS Properties}
  \label{tab:css-props}
  \begin{tabular}{l l}
    \toprule
    CSS Property & Example Values \\
    \midrule
    \textit{Timing} & \\
    \midrule
    duration & 5s \\
    timing-function & linear, ease-in-out \\
    delay & 1.5s \\ 
    \midrule
    \textit{Transforms} & \\
    \midrule
    scale, scaleX, scaleY & scale(1.5) \\
    translate, translateX, translateY &  translateY(-100\%) \\
    rotate, rotateX, rotateY & rotate(10deg) \\
    \midrule
    \textit{Other Appearance} & \\
    \midrule
    opacity & 0.5 \\ 
    fill & cyan, \#0000FF \\
    visibility & hidden, visible \\ 
    filter & brightness(100\%) \\
    font-family & Courier, monospace \\
    \midrule
    \textit{animation-} & \\
    \midrule
    direction & alternate, alternate-reverse, forward \\ 
    play-state & running \\
    fill-mode & both \\
  \bottomrule
\end{tabular}
\end{table}

\subsection{CSS Errors}
\label{app:css-errors}
We found 15 instances of syntactically incorrect generated CSS from GPT-4 out of the 233 total prompts entered during the study (a 6.7\% error rate) . Errors in CSS will still render the SVG within Keyframer, but the image will not appear animated. A summary of the errors identified are shown below. Note that we are using GPT-4 as is without any fine-tuning. Table \ref{tab:css-errors} summarizes the CSS errors identified.

\begin{table}[htb]
  \caption{CSS Syntax Errors}
  \label{tab:css-errors}
  \begin{tabular}{lc p{8cm}}
    \toprule
    Type & Count & Example \\
    \midrule
    Classname applied to keyframe & 6 (40.0\%) & \textbf{.design-9 @keyframes planet-rotate \{\}} instead of \textbf{@keyframes planet-rotate \{\}} \\
    \midrule
    Class notation instead of id & 3 (20.0\%) & \textbf{.design-0 .flame} instead of \textbf{.design-0 \#flame} \\
    \midrule
    Incorrect class refactoring & 2 (13.3\%) & \textbf{.design-0 \#rocketship \{\} .design-3 \#exhaust \{\}} (both should use \textbf{.design-3}) \\
    \midrule
    Typo & 2 (13.3\%) & \textbf{<stle>} instead of \textbf{<style>} \\
    \midrule
    Invalid CSS & 1 (6.6\%) & \textbf{\#sparkle-1, \#sparkle-2, \#sparkle-3 \{ animation-delay: 0.5s, 1s, 1.5s\}} \\
    \midrule
    Undefined variable & 1 (8\%) & \textbf{calc(30deg * var(—random))} when \textbf{random} has not been defined \\
  \bottomrule
\end{tabular}
\end{table}





\end{document}